\documentclass[a4paper,oneside,onecolumn,italian,12pt,draft]{article}
\usepackage{a4}
\usepackage{amsmath}
\usepackage{graphicx}

\input{tcilatex}

\begin{document}

\title{Work performed by a Classical-``reversible''-Carnot cycle:\\
Raising's distribution for the small ''driving weights''.}
\author{Francesco di Liberto\thanks{%
e-mail: diliberto@na.infn.it} \\
{\small \textit{INFM and INFN, Sezioni di Napoli}}\\
{\small \textit{Dipartimento di Scienze Fisiche, Universit\`{a} di Napoli
``Federico II''}} \\
{\small \textit{Complesso Universitario di Monte S. Angelo }}\\
{\small \textit{Via Cinthia \ Lotto G. \ \ I-80126 Napoli, Italy}}}
\maketitle

\begin{abstract}
The expansions or the compressions of the ideal gas in the quasi-static
Carnot cycle, can be performed (on adiabatic or isothermal way) by slowly
increasing or decreasing the external pressure by means of small weights
acting on the piston of the vessel containing the gas. We call them shortly
the ``driving weights'' ($dw$). Let $\mathit{N}$ be their number, a large
one.

To determine the work performed by the ideal gas in the cycle the ``driving
weights'' must be handled carefully. If we let them move on-off the piston
only horizontally, their vertical motions will be due only to the gas. Here
we show that, at the end, while some of them will have moved down (will have
negative raising) the remaining ones (the majority) will have moved up (will
have positive raising) so that the total work performed by the ideal gas \
equals the total variation of the gravitational potential energy of the
``driving weghts''.

The cycle is performed in $2N$ \ time-steps. For each step $t_{i},$ $i\in
(1,..,2N),$ we give $H(t_{i})\;$ and $\Delta H(t_{i-1},t_{i}),$ respectively
the height and the raising of the piston. Moreover the overall raising of
the individual $dw\prime s$ (i.e. $h_{k}$, $k\in $ $(1...\mathit{N)\;),}$
and their distribution are given in simple, general cases. The efficiency
and the dissipated work are also evaluated.

This paper is aimed at imparting a deeper understanding of the ideal Carnot
Engine and may also be useful as \ a segment in a didactic path on
elementary calculus and statistics.
\end{abstract}

\bigskip

PACKS: 05.50

\section{Introduction}

To perform an ideal gas quasi-static-``reversible''\cite{1}-Carnot-cycle, we
need an heat source, an heat sink, a vessel with a mobile piston and many
small ''driving weights'' to increase or decrease slowly the external
pressure (both for the isothermal and adiabatic processes).\cite{2}-\cite{5}

To determine the work performed by the ideal gas in the cycle the ``driving
weights'' ($dw$) must be handled carefully. Therefore we let them  move
on-off the piston only horizontally. To this end we assume that the handle
of the piston be endowed of so many consoles that we can move each $dw$
horizontally from (or to) the corresponding fixed console as reported in
Figure 1.

\FRAME{fpFU}{14.3879cm}{14.6537cm}{0pt}{\Qcb{a) The adiabatic vessel with
some $dw\prime s$ on the piston; $a=10\;cm,$ $b=27\;cm,$ $c=54\;cm.$ $%
V_{box}=abc=14580cm^{3}=14.58\;l.$ $V_{0}=$ $22.4l=V_{box}+$ $H_{0}\mathit{S,%
}$ where $\;H_{0}=78.2\;cm,\;\mathit{S\;}=aa=100\;cm^{2}$ b) The adiabatic
vessel together with two supports for the $dw\prime s$}}{\Qlb{1}}{%
figurone1-copia.jpg}{\special{language "Scientific Word";type
"GRAPHIC";display "PICT";valid_file "F";width 14.3879cm;height
14.6537cm;depth 0pt;original-width 4.7245in;original-height
7.0837in;cropleft "0.0854";croptop "0.8458";cropright "0.9847";cropbottom
"0.3063";filename '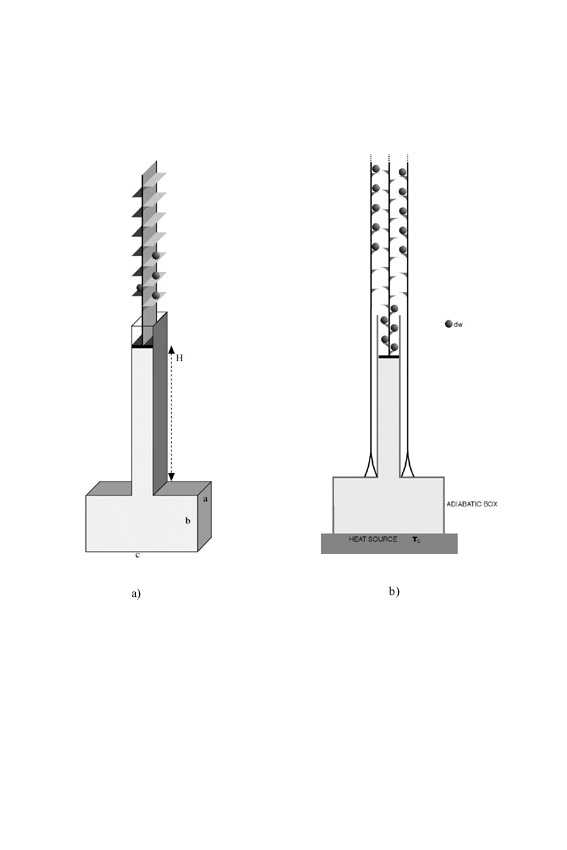';file-properties "XNPEU";}}The piston
is mobile, so when we increase the external pressure the last coming $dw$
(together with the previous ones) goes down. In the opposite case
(expansion) the chosen $dw$ goes at rest on the console in front of it and
the others go up.

``In this way you can find out the work performed by the ideal gas in the
cycle: you will find it in the increased potential energy of some of the $%
dw^{\prime }s$'', \ I said in the last 25 years in the thermodynamic class
at the University of Naples. Sometimes in these years some\ clever student
replayed:\ ``I have thought a lot about that and it seems to me that the $%
dw^{\prime }s$ go down''. \ My reply has been always ``You have to think
more, because the work performed by the gas must necessarily increase the
potential Energy of the $dw^{\prime }s$ \ ''.\cite{6}

Recently I decided to do a Java Applet \cite{7} to show the raising of the $%
dw^{\prime }s$ and,\ in writing down some elementary relations,\ I have
found out that the clever students were not totally wrong. In the following
I report on a simple Carnot-cycle done through the ideal device given in
Figure 1 and show that, at the end of the cycle, some of the $dw^{\prime }s$
will have moved down (will have negative raising) while the remaining ones
(the majority ) will have moved up (will have positive raising) so that the
total work performed by the ideal gas equals the total variation of the
gravitational potential energy of the ``driving weights''. In particular in
Sec.2 is given an example of a Carnot-cycle done in $2N$ steps. In the first 
$N$ steps the $dw^{\prime }s$ are added on the piston to perform first an
isothermal compression and then an adiabatic compression. In the remaining $N
$ steps the $dw^{\prime }s$ are removed from the piston in order to return
to the initial state. The cycle is reported in Figure 2. The height and the
raising of the piston et each step i.e. $H(t_{i})$ and $\ \Delta
H(t_{i-1},t_{i})=H(t_{i})-H(t_{i-1}),$ are also calculated and reported in
Figure 3 and Figure 4

In Sec.3 we evaluate the raising of each $dw$ at the end of the cycle. To
find the raising of the $dw$ which was put on the piston at one given time
step, one must know when it has been removed. The order in which the $%
dw^{\prime }s$ are removed is quite arbitrary and actually there are $N!$
ways in which the removing process can be done. We analyze two main
non-random removing processes: for both of them the overall raising of the
single $dw$ is related to the sum of the raisings of the piston for the
time-steps in which the $dw$ `lived' on the piston and hence to the
difference of the height of the piston between the time at which it was
removed and at which it was added on the piston . The results are shown in
Figure 6, Figure 7 and Figure 8. In Figure 5 one can understand, directly
from the $P-V$ representation of the cycle, why some $dw^{\prime }s$ have
negative raising. In Sect. 4 the relation which connects the raising of the
single $dw$ to the expansion of the gas is obtained in an alternative way.
The efficiency of the engine is evaluated in Sec.5 while the entropic gains
are estimated in Appendix.

\section{The cycle and the raising of the piston}

One mole of ideal gas (e.g. dry air) is contained in the vessel shown in
Figure 1, and\ it is initially in thermal contact with the heat reservoir at 
$T_{0}\,=273.15\;K$. The initial volume of the gas is $V_{0}=$ $22.4$ $\ l.$
Let $P_{0}=1\,\,atm=101.3\;KPa\;$be the external pressure on the piston and
let the piston be massless so that at the initial stage also the pressure of
the gas is $P_{0}.$ For the dry air $\gamma =\frac{C_{p}}{Cv}=1.4.$

The cycle we realize is reported in Figure 2. It is done in $2N$ time-steps.
In the first $N$ steps the pressure $P(t_{i}),$ $i\in (1..N)$ increases. At
each step we add on the piston a single $dw$. Let its mass be $m=0.1\;Kg.$
Since the surface of the piston is $\mathit{S\;}=100\;cm^{2},$ this implies
that at each step we are increasing the pressure of a relatively small
amount. The degree of irreversibility for this step-wise process\ is
evaluated in Appendix.

\FRAME{ftbpFU}{5.5953in}{4.2774in}{0pt}{\Qcb{The Carnot-cycle$.$ It is
step-wise but the steps are very small. $P_{0}V_{0}->P_{A}V_{A}$ is the
isothermal process at $T_{0}=273.15,\;($ $P_{A}=$ $1.3969$\ $atm$,\ $%
V_{A}=16.034$ $l).\;P_{A}V_{A}->P_{B}V_{B}$ is an adiabatic process\ ($%
P_{B}=1.5905$ $atm$\ and $V_{B}=14.614\;l$). $P_{B}V_{B}->P_{C}V_{C}\;$is the%
$\ $ isothermal process at $T=283.47\;(P_{C}=1.1385\ atm\ and\
V_{C}=20.416l).$ $P_{C}V_{C}->P_{0}V_{0}$ is the final adiabatic process.}}{%
\Qlb{2}}{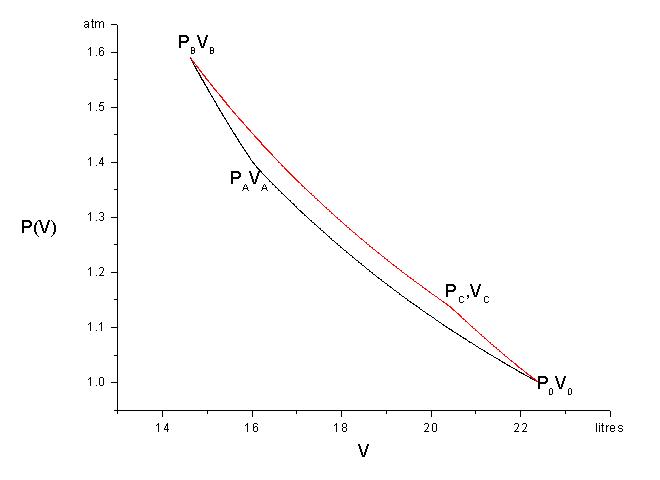}{\special{language "Scientific Word";type
"GRAPHIC";maintain-aspect-ratio TRUE;display "USEDEF";valid_file "F";width
5.5953in;height 4.2774in;depth 0pt;original-width 27.8418in;original-height
21.2502in;cropleft "0";croptop "1";cropright "1";cropbottom "0";filename
'fig2.jpg';file-properties "XNPEU";}}

In the first $N$ steps we have 
\begin{equation*}
\frac{\;P(t_{i})}{P_{0}}=\frac{z+i}{z}
\end{equation*}
where $P(t_{0})\equiv P_{0}\;$is the initial pressure, $i$ is the number of $%
dw^{\prime }s$ on the piston at step$\;t_{i}$ and $\ z=1033$ is the number
of grams whose weight on $1\;cm^{2}$ give the pressure of $1\,\,atm.$
Therefore $P(t_{i})=P_{0}+i\Delta P$ with $\Delta P=\frac{1}{z}P_{0}=\frac{1%
}{1033}P_{0}.$ In the second $N$ steps the $dw^{\prime }s$ are removed. As
at each step a single $dw$ is removed, for $P(t_{N+l}),$ $l\in (1...N)$ we
have $P(t_{N+l})=P_{0}+(N-l)\Delta P$ and 
\begin{equation*}
\frac{P(t_{N+l})}{P_{0}}=\frac{z+N-l}{z}=\frac{P(t_{N-l})}{P_{0}}
\end{equation*}
\ but obviously $V(t_{N+l})\neq V(t_{N-l})$ .

The choice of the values of $N$ and $n_{1}$ (the number of steps of the
isothermal process at $T_{0})$ are somehow free, even if they depend on the
geometry of the vessel and on \ choice of the mass of the single $dw$. Here
we find convenient the values $N=610$ \ and $n_{1}=410$.

The state $P_{A}V_{A}$ is therefore reached in $n_{1}$ time-steps in
isothermal conditions. The height of the piston at each step, i.e. $H(t_{i})$
can be determined from the\ ideal gas state equation $%
P(t_{i})V(t_{i})=RT_{0},$ which, together with the relation 
\begin{equation}
V(t_{i})=V_{box}+H(t_{i})\mathit{S}
\end{equation}
gives 
\begin{equation}
H(t_{i})=H_{0}+\frac{V_{0}}{\mathit{S}}(\frac{P_{0}}{P(t_{i})}-1)
\end{equation}
From this we can evaluate step by step the raising of the piston $\ \Delta
H(t_{i-1},t_{i})=H(t_{i})-H(t_{i-1})\equiv \Delta H(t_{i}),$ therefore

\begin{equation}
\Delta H(t_{i})=\frac{V_{0}P_{0}}{\mathit{S}}\left( \frac{1}{P(t_{i})}-\frac{%
1}{P(t_{i-1})}\right)
\end{equation}
Let denote \ with $H_{1}(t_{i})$ and $\Delta H_{1}(t_{i})$ the values given
by relations (2) and (3), which are relative to the first $n_{1}$ steps.

It is obvious that for each isothermal step $\Delta T(t_{i})=0;$ not obvious
is the estimate of the Entropic change of the Universe for an isothermal
step, in the Appendix we find $\Delta\mathcal{S}_{U}(t_{i})\simeq$ $R(\frac{%
\Delta P}{P})^{2},$ where $R$ is the Universal gas constant.

The state $P_{B}V_{B}$ \ is reached in $n_{2}=$ 200 time-steps in adiabatic
conditions (thermal contact replaced by an adiabatic wall). For the
adiabatic steps\textit{\ }we cannot use $P(t_{i})V(t_{i})^{\gamma }=$ $%
Const. $ Nevertheless by means of the First Law \ of thermodynamics and the
equation of state of the ideal gases in the Appendix we find that

\begin{equation*}
\Delta V(t_{i})=-\frac{V(t_{i-1})}{\gamma }\frac{\Delta P}{P(t_{i})}
\end{equation*}
i.e. 
\begin{equation}
\Delta H(t_{i})=-\frac{V(t_{i-1})}{\gamma \mathit{S}}\frac{\Delta P}{P(t_{i})%
}
\end{equation}
Let call $\Delta H_{2}(t_{i})$ the raising of piston. From (4) we can
evaluate $H_{2}(t_{i}),$ the height of the piston in each of the $n_{2}$
adiabatic steps.

The final height$\;$reached is $H_{B}=H_{2}(t_{N})\ =0.344\mathit{\ cm.}$%
\textit{\ }

\FRAME{ftbpFU}{5.4068in}{3.3122in}{0pt}{\Qcb{Height of the piston step by
step.$H_{0}=78.2cm.$ $H_{B}=0.342cm$}}{\Qlb{3}}{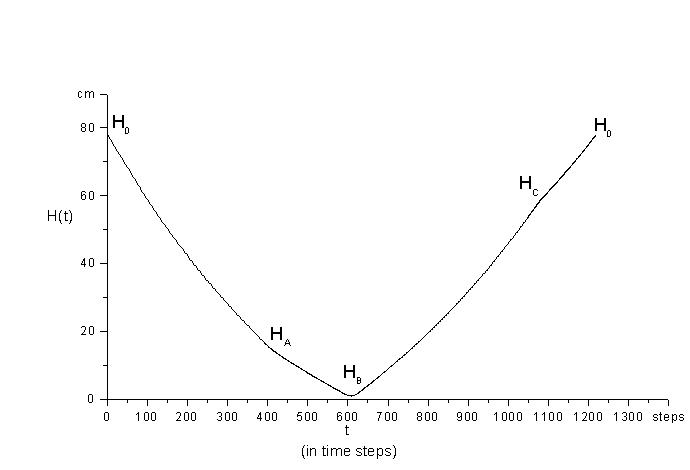}{\special{ language
"Scientific Word"; type "GRAPHIC"; maintain-aspect-ratio TRUE; display
"USEDEF"; valid_file "F"; width 5.4068in; height 3.3122in; depth 0pt;
original-width 30.0669in; original-height 18.3591in; cropleft "0"; croptop
"1"; cropright "1"; cropbottom "0"; filename 'fig3.jpg';file-properties
"XNPEU";}}

In the adiabatic compression the temperature of the ideal gas increases at
each time-step. The\ thermal increase at each step can\ be calculated taking
in account the First Law of Thermodynamics for an adiabatic process, i.e. $%
\Delta U=-P\Delta V,$ and the ideal gas property $\Delta U=C_{V}\Delta T,$
where $U$ is the Internal Energy and $C_{V}$ the molar specific heat at
constant volume. We have 
\begin{equation}
\Delta T(t_{i})=-\frac{P(t_{i})}{C_{V}}\mathit{S}\Delta H(t_{i})
\end{equation}
The final temperature is $T_{B}=283.47K^{o}.$ The entropic change for an
adiabatic step is $\Delta \mathcal{S}_{U}(t_{i})\simeq \frac{R}{\gamma }(%
\frac{\Delta P}{P})^{2}\;\;($see Appendix). To reach the initial state $%
P_{0}V_{0},$ we need $n_{3}=467$ isothermal steps and $n_{4}=143$ adiabatic
steps.

Removing $n_{3}$ $dw^{\prime }s$ from the piston in isothermal conditions
(thermal contact with the heat source $T_{B})$ we get the state $P_{C}V_{C}.$
\ For this expansion we have 
\begin{equation}
H_{3}(t_{i})=\frac{P_{B}V_{B}}{\mathit{S}P(t_{i})}-\frac{V_{box}}{\mathit{S}}%
=H_{B}+\frac{V_{B}}{\mathit{S}}(\frac{P_{B}}{P(t_{i})}-1)
\end{equation}
from which the raisings of the piston $\Delta H_{3}(t_{i})$ can be
evaluated. Observe that in the expansions $\Delta H\geq 0\ \ \ $and$\ \frac{%
\Delta P}{P_{0}}=-\frac{1}{1033}.$ Moreover $\Delta T(t_{i})=0\;\;\;$and $\
\ \Delta \mathcal{S}(t_{i})\simeq R(\frac{\Delta P}{P})^{2}$ (see Appendix)

Finally removing the last $n_{4}$ $dw^{\prime }s$ in adiabatic conditions we
come back to the initial state\ $V_{0},P_{0},T_{0}.$ In this last process we
can evaluate $\Delta H_{4}(t_{i})$ trough a relation similar to (4). $%
H_{4}(t_{i})$ can consequently be determined. Now at each step the
temperature decreases; the entropic change, as before, is $\Delta \mathcal{S}%
_{U}(t_{i})\simeq \frac{R}{\gamma }(\frac{\Delta P}{P})^{2}\;\;($see
Appendix)

In Figure 3 we give the height of the piston for each step. Fig 4 reports
the raising of the piston for each step.\FRAME{ftbpFU}{4.8369in}{3.3122in}{%
0pt}{\Qcb{Raising of the piston step by step.Observe the discontinuities in
the raising of the piston going from the isothermal process to the adiabatic
process and \textit{vice-versa.} i.e. around the states A, B and C}}{\Qlb{4}%
}{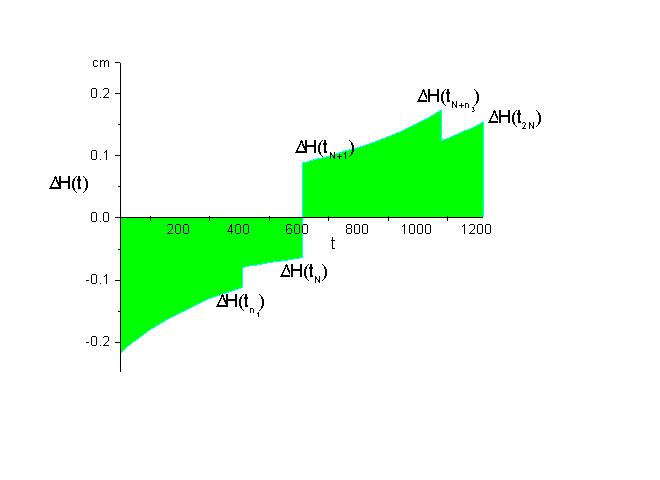}{\special{language "Scientific Word";type
"GRAPHIC";maintain-aspect-ratio TRUE;display "USEDEF";valid_file "F";width
4.8369in;height 3.3122in;depth 0pt;original-width 23.2911in;original-height
15.9082in;cropleft "0";croptop "1";cropright "1";cropbottom "0";filename
'fig4.jpg';file-properties "XNPEU";}}

\section{\protect\bigskip Raising of the single $dw^{\prime }s$}

In this section we find the raising of each $dw$ and show that some of them
move down (have negative raising) and that the remaining majority move up
(have positive raising). The history of the single $dw$ is relevant to
evaluate its raising. To this end let us label each $dw$. The $k^{th}$ $dw$
is the one placed on the piston at the $k^{th}$ time-step ($k\in (1...N))$.
So the $N^{th}dw$ is the last one. The raising of the $k^{th}$ $dw$ at the
end of the cycle, $h_{k},$ is its vertical shift on the support, i.e. the
difference between the final and the initial position on the supports.

Really relevant is the order in which they are removed from the piston. Two
are the possible ways in which we can start to remove them: $a)$ from the $%
N^{th},\;b)$ from the $(N-L)^{th}\;\ $with \ $1\leq L<N$.

In the following we report on both cases for non-random processes: \ in the
case $a)$ we start from the $N^{th}$ and continue with the ($N-1)^{th}$
until the $1^{th};$ in the case $b)$ we start from the $(N-L)^{th},$ go on
with the $(N-L+1)^{th}$ until the $N^{th}$ and then continue with the $%
(N-L-1)^{th}$ until the $1^{th}.\;$We call these processes respectively $%
a-processes$ and $b-processes$. Obviously the expansion process can be done
in $N!$ ways. For example you can start with $N^{th}\;dw,$ go regularly to
the $(N-R)^{th}$ then jump to the $(N-2R)^{th}$ and then return to the $%
(N-R)^{th}$ and continue; otherwise you can start from the $(N-L)^{th}$ go
to the $N^{th},$ jump to the $(N-2L)^{th},$ return to the $(N-L-1)^{th}$ and
so on and so on. Once all the possible regular processes have been exhausted
one can go to the random-processes. The nice aspect is that for each process
we have a different distribution of the $h_{k}$. With the complexity of the
distribution of the $h_{k}$ we recover some of the complexity of the
microscopic behavior of the ideal gas.

$a-processes.$ Let the first removed be the $N^{th}.$ The individual raising
are related to the raising of the piston, therefore it is clear that $h_{N}$%
, the raising of the last $dw$ (which has been on the piston just for one
step) is given by $h_{N}=\Delta H(t_{N})=H(t_{N})-H(t_{N-1})$ and for the
last but one $dw$ (which has been on the piston just for two steps) it is
clear that $h_{N-1}=\Delta H(t_{N-1})+\Delta H(t_{N})+\Delta
H(t_{N+1})=H(t_{N+1})-H(t_{(N-1)-1}).$

Therefore for the $(N-r)^{th}$ \ $dw$, with $r\in (0,..,N-1),$ we have 
\begin{align}
h_{N-r}& =\sum_{\;i=N-r}^{N+r}\Delta H(t_{i})=\sum_{\;i=N-r}^{N}\Delta
H(t_{i})+\sum_{\;i=N+1}^{N+r}\Delta H(t_{i})= \\
& =H(t_{N})-H(t_{N-r-1})+H(t_{N+r})-H(t_{N+1-1})=  \notag \\
& =H(t_{N+r})-H(t_{N-r-1})  \notag
\end{align}
which immediately gives 
\begin{equation}
h_{N-r}=\frac{1}{\mathit{S}}[\;V(t_{N+r})-V(t_{N-r-1})\;]
\end{equation}
This last relation is useful since it enables to appreciate the raising of
the $dw$ directly from an inspection of the $P-V$ diagram of the
Carnot-cycle : we have only to observe that $V(t_{N+r})$ is the volume
occupied by the gas in the \textit{expansion} at pressure $P(t_{N+r})$ and
that $V(t_{N-r-1})$ is the volume in the \textit{compression }at the lower
pressure $P(t_{N-r-1})=P(t_{N-r})-\Delta P.$ In this way at each step we can
find $\delta V=V(t_{N+r})-V(t_{N-r-1})$, which is relative to just one $%
\Delta P.$ From the $P-V$ diagram of the Carnot-cycle we see that $\delta V$
is negative only for the first steps and last steps, respectively. In fig 5,
on a schematic representation of the extremities of the cycle, some positive
and negative $\delta V$'s are shown. Therefore we see that the raising $h_{k}
$ is \textit{negative} only for the first and the last steps and is \textit{%
positive} for all the others. From this analysis, moreover, we can
understand that for the $a-processes$ the number of negative raisings
depends on how big is $\Delta P.$ For $\Delta P\rightarrow 0$ that number
goes to zero.

\FRAME{ftbpFU}{5.2702in}{2.3955in}{0pt}{\Qcb{a) Positive and negative values
of $\Delta V$ around the $N^{th}$ step. b) Positive and negative values of $%
\Delta V$ around the $2N^{th}$ step. The representation is schematic: the
step-wise aspect of the cycle is not reported .}}{\Qlb{5}}{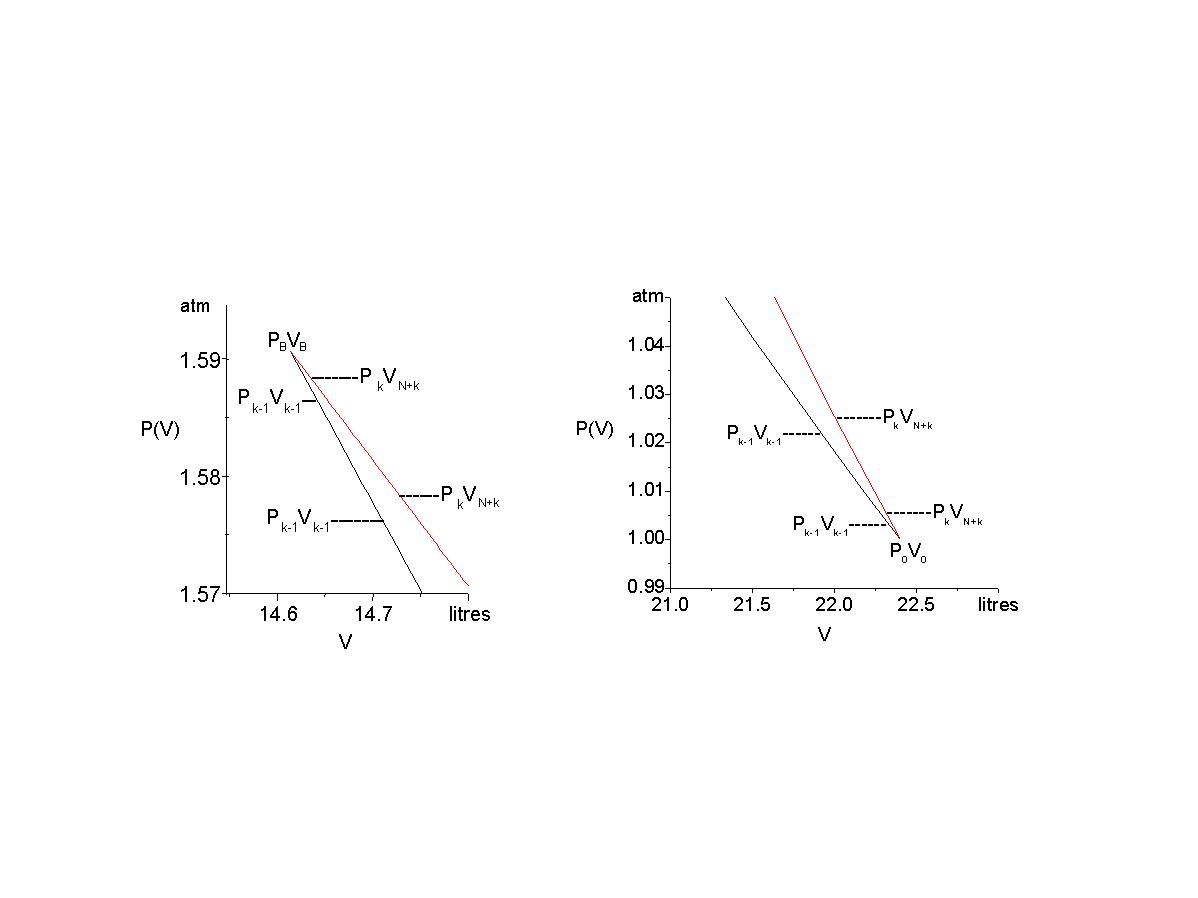}{\special%
{language "Scientific Word";type "GRAPHIC";maintain-aspect-ratio
TRUE;display "USEDEF";valid_file "F";width 5.2702in;height 2.3955in;depth
0pt;original-width 9.9912in;original-height 7.491in;cropleft
"0.0963";croptop "0.7430";cropright "0.9036";cropbottom "0.2567";filename
'fig5.jpg';file-properties "XNPEU";}}

The values of \ $h_{N-r}$ are obtained from equation (8) together with
relations (2), (4), (6) and are reported in Figure 6.

\FRAME{ftbpFU}{4.8966in}{5.5175in}{0pt}{\Qcb{Distribution of the overall
rising for each dw in the $a-processes$. The overall rising is the vertical
shift of the single dw on the support i.e. the difference between the final
position and the initial position on the support. The top two graphs are
magnifications of the initial and final part of the bottom graph and show
that the last dw's and the first ones have negative raising}}{\Qlb{6}}{%
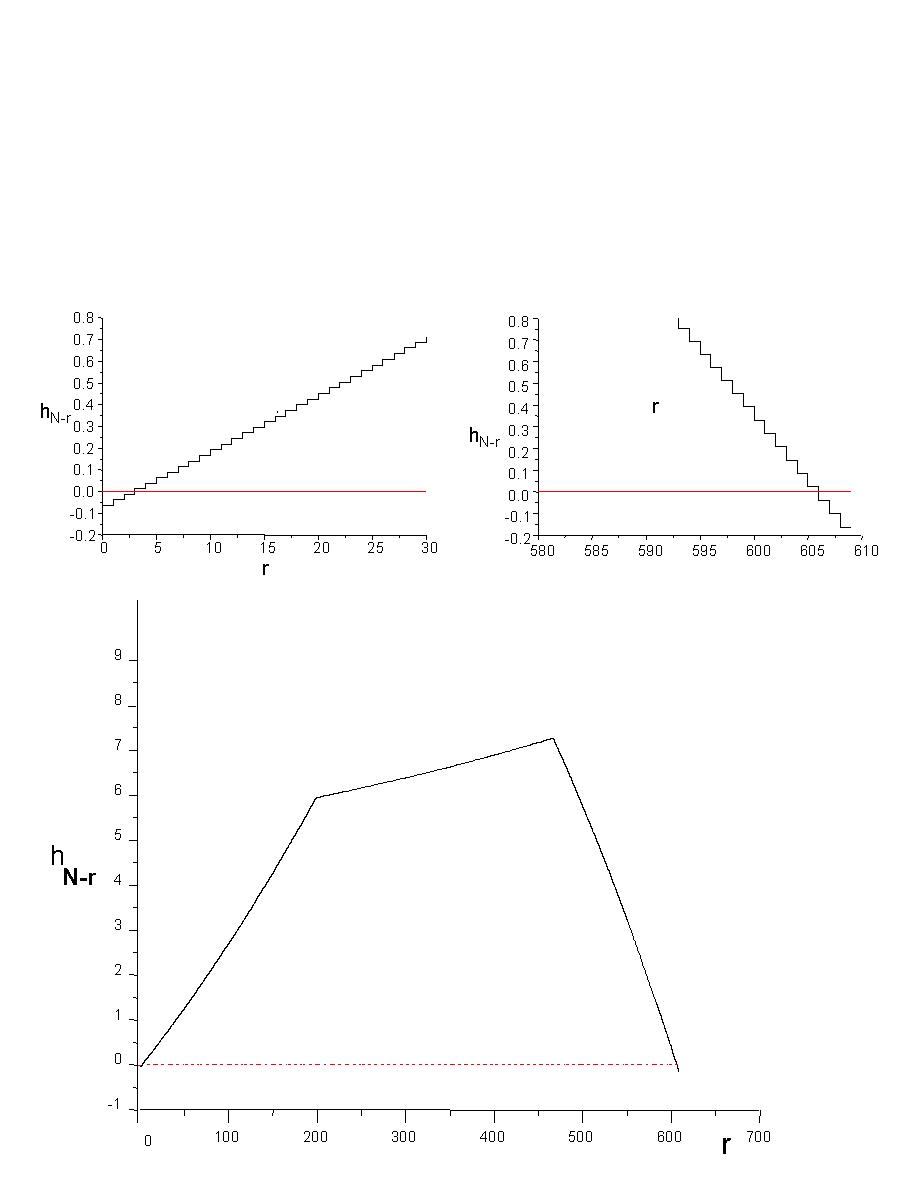}{\special{language "Scientific Word";type
"GRAPHIC";maintain-aspect-ratio TRUE;display "FULL";valid_file "F";width
4.8966in;height 5.5175in;depth 0pt;original-width 7.491in;original-height
9.9912in;cropleft "0";croptop "0.8454";cropright "1";cropbottom "0";filename
'fig6.jpg';file-properties "XNPEU";}}

Now we go to $b-processes.$

If the first $dw$ removed is the $\ (N-L)^{th}\ $with \ $1\leq L<N$, the
histories\ of the single $dw^{\prime }s$ change. The $(N-L)^{th}dw$ has been
on the piston during the last $L+1$ steps of the compression and we have 
\begin{equation}
h_{N-L}=\sum_{i=N-L}^{N}\Delta H(t_{i})=H(t_{N})-H(t_{N-L-1}).
\end{equation}
For $\left( N-(L-1)\right) ^{th}dw$, which has been on the piston for the
last $L$ steps of the compression and for the first step of the expansion
(that one in which the $(N-L)^{th}$ $dw$ is removed) we have 
\begin{equation}
h_{N-(L-1)}=\sum_{\;i=N-(L-1)}^{N}\Delta H(t_{i})+\Delta
H(t_{N+1})=H(t_{N+1})-H(t_{N-L})
\end{equation}
Therefore, if we denote with $h_{N-r}^{L}\;$the\textsl{\ }$b-processes$
raising, for $r\leq L,$ we have 
\begin{equation}
h_{N-r}^{L}=\sum_{\;i=N-r}^{N}\Delta H(t_{i})+\sum_{\;i=N+1}^{N+L-r}\Delta
H(t_{i})=H(t_{N+L-r})-H(t_{N-r-1})
\end{equation}
which gives 
\begin{equation}
h_{N-r}^{L}=\frac{1}{\mathit{S}}[V(t_{N+L-r})-V(t_{N-r-1})]
\end{equation}
This relation too is useful to appreciate the raising of the $dw^{\prime }s$
directly from an inspection of the $P-V$ diagram of the Carnot-cycle : we
can observe that $V(t_{N+L-r})$ is the volume occupied by the gas in the 
\textit{expansion} at pressure $P(t_{N+L-r})=P_{0}(z+N-(L-r))/z$ and that $%
V(t_{N-r-1})$ is the volume in the \textit{compression }at the pressure $%
P(t_{N-r-1})=P_{0}(z+N-r-1)/z.$ So $\delta V=$ $V(t_{N+L-r})-V(t_{N-r-1})$ \
is relative to a $\delta P=P_{0}(2r+1-L)/z,$ and it is possible to see that
for $\delta P>0$ we have $\delta V<0$ and hence $h_{N-r}^{L}<0$ and for $%
\delta P<0$ we have $\delta V>0$ and hence $h_{N-r}^{L}>0$. For example, for 
$r=L$ \ we have $\delta P=$\ $\delta P_{\max }=\frac{L+1}{z}P_{0}$ $>0$, to
which corresponds the maximum negative raising $h_{N-L}^{L};$ for $r=0$ \ we
have $\delta P=$\ $\delta P_{\min }=\frac{1-L}{z}P_{0}<0$,\ \ to which
corresponds the maximum positive raising $h_{N}^{L}$. The values of $%
h_{N-r}^{L}$ can be calculated using relation (12) together with (2), (4),
(6).

For $r>L,$ the way in which the previous $L$ $dw^{\prime }s$ have been
removed has no influence. Therefore, for $r>L$ everything is as in the $%
a-processes$ 
\begin{equation}
h_{N-r}^{L}=h_{N-r}=H(t_{N+r})-H(t_{N-r-1})
\end{equation}
In Fig 7 and Fig 8\ we report the distributions of $h_{N-r}^{L}$ for $L=25$
and $L=50$.

\FRAME{ftbpFU}{4.6432in}{3.3122in}{0pt}{\Qcb{Distribution of the overall
rising for each dw in the $b-processes$ for $L=25$}}{\Qlb{7}}{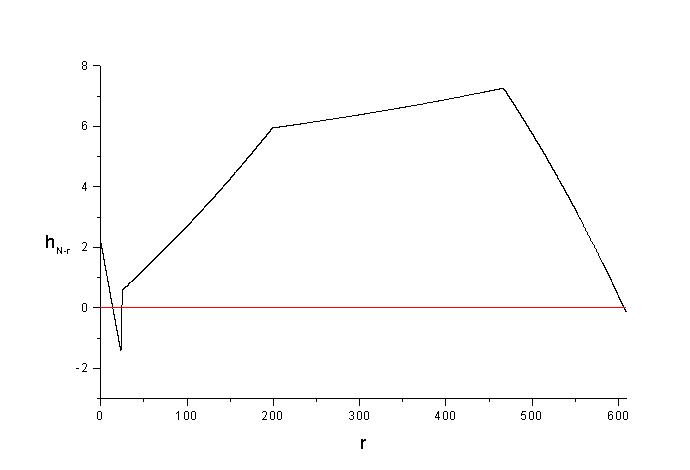}{%
\special{language "Scientific Word";type "GRAPHIC";maintain-aspect-ratio
TRUE;display "USEDEF";valid_file "F";width 4.6432in;height 3.3122in;depth
0pt;original-width 27.0998in;original-height 19.287in;cropleft "0";croptop
"1";cropright "1";cropbottom "0";filename 'fig7.jpg';file-properties
"XNPEU";}}

\FRAME{ftbpFU}{5.2036in}{3.3122in}{0pt}{\Qcb{Distribution of the overall
rising for each dw in the $b-processes$ for $L=50.$}}{\Qlb{8}}{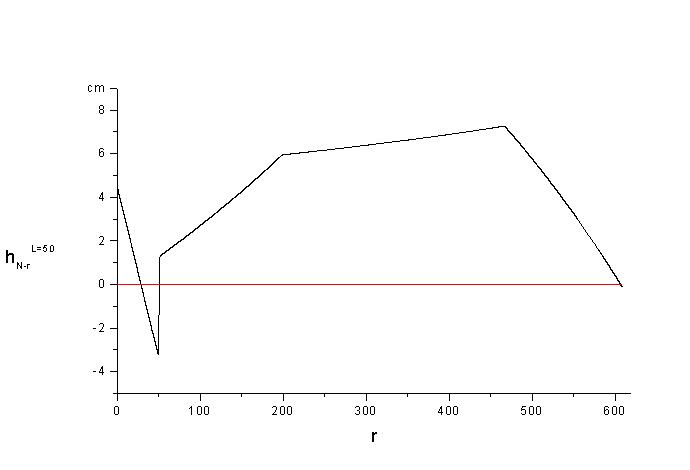}{%
\special{language "Scientific Word";type "GRAPHIC";maintain-aspect-ratio
TRUE;display "USEDEF";valid_file "F";width 5.2036in;height 3.3122in;depth
0pt;original-width 28.5155in;original-height 18.0953in;cropleft "0";croptop
"1";cropright "1";cropbottom "0";filename 'fig8.jpg';file-properties
"XNPEU";}}We point out that the relationship 
\begin{equation}
\sum_{r=0}^{L}h_{N-r}^{L}=\sum_{r=0}^{L}h_{N-r}
\end{equation}
is fulfilled for these two processes since the work performed by the ideal
gas in the cycle is the same, whichever removing process is performed. This
identity can be verified using relations (8) and (12).

We conclude this section pointing out that a deeper insight in the cycle can
be obtained through the time-dependent raising $h_{N-r}(t_{N+l})$ for $l\in
(1...N).$ The raising we are dealing with in this paper are the $%
h_{N-r}(t_{2N})$ i.e. the raising of the $dw^{\prime }s$ \textit{at the end }%
of the cycle.

\section{Engine work and raising of the $dw^{\prime }s$}

The work performed by the ideal gas in our step-wise Carnot-cycle, $W,$ is
clearly given by 
\begin{equation}
W=\sum_{i=1}^{2N}P_{i}\Delta V_{i}
\end{equation}
For a reversible cycle 
\begin{equation}
W_{rev}=\oint PdV
\end{equation}
It comes from the elementary calculus that it can be written. 
\begin{equation}
\oint PdV-\sum_{i=1}^{2N}P_{i}\Delta V_{i}\simeq \frac{\left| \Delta
P\right| }{P}.
\end{equation}
From physics point of view it is clear that

\begin{equation}
\sum_{i=1}^{2N}P_{i}\Delta V_{i}=m\mathbf{g}\sum_{i=1}^{N}h_{i}
\end{equation}
where\textbf{\ }$\mathbf{g}$ \textbf{\ }is the gravity acceleration; but to
prove this relation we need a little of algebra. It is worth while to write
down the proof of this relation in order to have an alternative deduction of
the relation we have found for $h_{i}.$

Observe that using the identity$\sum_{i=1}^{2N}P_{0}\Delta V_{i}=0$ we can
write $\sum_{i=1}^{2N}P_{i}\Delta V_{i}$ in the following way 
\begin{equation}
\sum_{i=1}^{2N}P_{i}\Delta V_{i}=\Delta P\sum_{i=1}^{N}i\;\Delta
V(t_{i})+\Delta P\sum_{l=1}^{N}(N-l)\Delta V(t_{N+l})
\end{equation}
And from $\Delta V(t_{j})=V(t_{j})-V(t_{j-1})$ and the identity$%
\sum_{i=0}^{N-1}V(t_{i})=\sum_{l=0}^{N-1}V(t_{N-l-1})$ we have 
\begin{equation}
\sum_{i=1}^{2N}P_{i}\Delta V_{i}=\Delta P\sum_{l=0}^{N-1}\left(
V(t_{N+l})-V(t_{N-l-1})\right)
\end{equation}
Now recalling that 
\begin{equation*}
\Delta P=\frac{1}{z}P_{0}=\frac{1}{z}\frac{z\;g}{cm^{2}}\mathbf{g}=\frac{m}{%
\mathit{S}}\mathbf{g}
\end{equation*}
we can conclude that 
\begin{equation}
\sum_{i=1}^{2N}P_{i}\Delta V_{i}=\frac{m}{\mathit{S}}\mathbf{g}%
\sum_{l=0}^{N-1}[V(t_{N+l})-V(t_{N-l-1})]=m\mathbf{g}%
\sum_{l=0}^{N-1}h_{N-l}=m\mathbf{g}\sum_{i=1}^{N}h_{i}
\end{equation}
with 
\begin{equation}
h_{N-l}=\frac{1}{\mathit{S}}[V(t_{N+l})-V(t_{N-l-1})]
\end{equation}
This relation together with the equality (18) can give property (8) in an
alternative way.

\section{Efficiency of the Engine}

The heat quantity $Q_{a}\;$that the engine adsorbs in the $n_{3}\;$steps
performed in thermal contact with the heat reservoir at $T_{B}$ is given by 
\begin{equation}
Q_{a}=\sum_{i=1}^{n_{3}}P(t_{N+i})\Delta V(t_{N+i})
\end{equation}
The efficiency of the engine \ is therefore given by 
\begin{equation}
\eta =\frac{\sum_{i=1}^{2N}P_{i}\Delta V_{i}}{\sum_{i=1}^{n_{3}}P(t_{N+i})%
\Delta V(t_{N+i})}
\end{equation}
As is well known, the adiabatic works in the Carnot-cycle cancel each other,
therefore 
\begin{equation}
\eta =1+\frac{\sum_{i=1}^{n_{1}}P(t_{i})\Delta V(t_{i})}{%
\sum_{i=1}^{n_{3}}P(t_{N+i})\Delta V(t_{N+i})}
\end{equation}
of course 
\begin{equation*}
\sum_{i=1}^{n_{1}}P(t_{i})\Delta V(t_{i})=Q_{0}
\end{equation*}
where $Q_{0}\;$is the heat delivered at the heat reservoir $T_{0}.$ From
relation (29) in the Appendix we know that in the isothermal compression $%
\frac{-\Delta V}{V(t_{i})}=\frac{\Delta P}{P(t_{i-1})},$ therefore 
\begin{equation}
Q_{0}=\sum_{i=1}^{n_{1}}P(t_{i})V(t_{i})\;\left[ \frac{\Delta V(t_{i})}{%
V(t_{i})}\right] =-RT_{0}\sum_{i=1}^{n_{1}}\frac{\Delta P}{P(t_{i-1})}%
=-RT_{0}\sum_{i=1}^{n_{1}}\frac{1}{z+i-1}
\end{equation}
The sum is 
\begin{equation*}
\sum_{i=1}^{n_{1}}\frac{1}{z+i-1}=\frac{\partial }{\partial z}\left[ \ln
(z-1+n_{1})!-\ln (z-1)!\right] =\Psi (z-1+n_{1})-\Psi (z-1)
\end{equation*}
where $\Psi (z)$ is usually called the `digamma function', i. e. the
logaritmic derivative of the $\Gamma (z)$ function.

For $Q_{a}$ we can similarly write 
\begin{equation}
Q_{a}=\sum_{i=1}^{n_{3}}P(t_{N+i})\Delta V(t_{N+i})=RT_{B}\sum_{i=1}^{n_{3}}%
\frac{1}{z+N-n_{3}-1+i-1}
\end{equation}
from which the efficiency is 
\begin{equation}
\eta =1-\frac{T_{0}}{T_{B}}f(z,n_{1},n_{3},N)
\end{equation}
where 
\begin{equation*}
f(z,n_{1},n_{3},N)=\frac{\Psi (z-1+n_{1})-\Psi (z-1)}{\Psi (z-1+N-1)-\Psi
(z-1+N-n_{3}-1)}
\end{equation*}
in our example ($z=1033,n_{1}=410,n_{3}=467,N=610)$%
\begin{equation*}
\eta =1-\frac{T_{0}}{T_{B}}(1+\epsilon )
\end{equation*}
with $\epsilon =1.2\;\;10^{-3}.$ We can therefore conclude that the
efficiency of our step-wise ideal engine is smaller that of the
corresponding reversible Carnot engine. A result which was expected \ since
the ``Dissipated Work''$\;\;W_{D}$ (see Appendix)\ is positive 
\begin{equation}
W_{D}=\sum_{i=1}^{2N}T(t_{i})\Delta \mathcal{S}_{U}(t_{i})>0  \notag
\end{equation}
Obviously it is also expected that for $\Delta P->0$ (i.e. $N->\infty )$ $\
\eta =1-\frac{T_{0}}{T_{B}}$

\section{Conclusions}

The detailed analysis of the classical Carnot cycle we have followed here
shows a fruitful complexity in the behaviour of small driving weights, whose
energetic gain can stimulate further speculations. A deeper insight in the
cycle can be obtained trough the time-dependent raisings $h_{N-r}(t_{i}),$
for instance it would be interesting to study the effect, on the individual
raising $h_{N-r}(t_{i}),\;$of the fact that the adiabatic works in the
Carnot-cycle for an ideal gas cancel each other. The estimate of the
Efficiency and of the Dissipated Work in our step-wise cycle may be useful
to give a deeper insight on the relation among them. In particular we plan
to show in a forthcoming paper that the efficiency of an arbitrary non
reversible engine running between $T_{\min }$ and $T_{\max }$, $\eta
_{Irre}(T_{\min },T_{\max }),$ is equal to the efficiency of a suitable
step-wise ideal gas Carnot engine $\eta _{step}(T_{\min },T_{\max },\Delta
P).$ All these are conceptual aspects. It would moreover  be useful to have
some practical realization of the actual step-wise Carnot engine since the
pattern of the overall raising of the $dw^{\prime }s$ would be preserved in
spite of the energy loss due to the friction between the piston and the
vessel.

\appendix

\section{Entropic changes for Isothermal and Adiabatic Processes of the
step-wise Carnot-cycle}

1$)$ Isothermal processes

For each time-step $t_{i}$ we have $P(t_{i})V(t_{i})=RT,$ $T$ being the
constant temperature at which the process is performed

To evaluate $\Delta \mathcal{S}_{U}(t_{i})=\Delta \mathcal{S}%
_{sys}(t_{i})+\Delta \mathcal{S}_{env}(t_{i})$ we first observe that :

\begin{equation*}
\ \Delta U=0\;\Rightarrow \Delta Q=\Delta W=P\Delta
V=P(t_{i})[V(t_{i})-V(t_{i-1})]
\end{equation*}
and 
\begin{equation}
\frac{V(t_{i-1})-V(t_{i})}{V(t_{i})}=\frac{P(t_{i})-P(t_{i-1})}{P(t_{i-1})}%
\Longrightarrow \frac{-\Delta V}{V}=\frac{\Delta P}{P}.
\end{equation}
Note that during the expansion $V(t_{i-1})-V(t_{i})$ \TEXTsymbol{>} 0
whereas during the compression $V(t_{i-1})-V(t_{i})<$ 0, so in the following
we will use $\left| \Delta V\right| ,$ when necessary. So

\begin{align}
\Delta \mathcal{S}_{sys}(t_{i})& =\mathcal{S}_{sys}(t_{i})-\mathcal{S}%
_{sys}(t_{i-1})=\int_{t_{i-1}}^{t_{i}}\frac{\delta Q}{T}%
=\int_{t_{i-1}}^{t_{i}}\frac{PdV}{T}=  \notag \\
& =R\;\ln \frac{V(t_{i})}{V(t_{i-1})}=R\;\ln \left( 1-\frac{%
V(t_{i-1})-V(t_{i})}{V(t_{i-1})}\right) \cong   \notag \\
& \cong R\left[ \left( -\frac{V(t_{i-1})-V(t_{i})}{V(t_{i-1})}\right) +\frac{%
1}{2}\left( \frac{V(t_{i-1})-V(t_{i})}{V(t_{i-1})}\right) ^{2}+...\right] 
\end{align}
and 
\begin{equation}
\Delta \mathcal{S}_{env}(t_{i})=\mathcal{S}_{env}(t_{i})-\mathcal{S}%
_{env}(t_{i-1})=\frac{\Delta Q}{T}=R\left( \frac{V(t_{i-1})-V(t_{i})}{%
V(t_{i})}\right) 
\end{equation}
Therefore to the first order, using relations (30) and (31) we have 
\begin{equation}
\Delta \mathcal{S}_{U}(t_{i})=R\left[ \frac{\left| \Delta V\right| ^{2}}{%
V(t_{i-1})V(t_{i})}\right] \cong R\left( \frac{\left| \Delta V\right| }{V}%
\right) ^{2}
\end{equation}
and, using (29), the entropic gain in an isothermal step is 
\begin{equation}
\Delta \mathcal{S}_{U}(t_{i})\cong R\left( \frac{\Delta P}{P}\right) ^{2}
\end{equation}
2) Adiabatic processes

For the adiabatic steps \textit{a priori }we cannot use $P(t_{i})V(t_{i})^{%
\gamma }=$ $C.$ Nevertheless for an adiabatic compression-step, from the
First Law of thermodynamics we have$\ \Delta U=-\Delta W=-P\Delta
V=P(t_{i})[V(t_{i-1})-V(t_{i})]>0,$ but $\Delta U=C_{V}(T(t_{i})-T(t_{i-1})),
$ so from the state equation $PV=RT$ we obtain 
\begin{equation*}
C_{V}(T(t_{i})-T(t_{i-1}))=R[\frac{P(t_{i})}{P(t_{i-1})}T(t_{i-1})-T(t_{i})]
\end{equation*}
i.e. 
\begin{equation}
(C_{V}+R)T(t_{i})=RT(t_{i-1})\left( \frac{P(t_{i})}{P(t_{i-1})}+\frac{C_{V}}{%
R}\right) 
\end{equation}
which gives \ $T(t_{i})$ and therefore, since $C_{P}=C_{V}+R$%
\begin{equation*}
V(t_{i})=\frac{RP(t_{i-1})V(t_{i-1})}{P(t_{i})C_{P}}\left( \frac{P(t_{i})}{%
P(t_{i-1})}+\frac{C_{V}}{R}\right) 
\end{equation*}
and recalling that$\;\ \gamma =c_{P}/c_{V}$ \ this can be written 
\begin{align}
\frac{V(t_{i})}{V(t_{i-1})}& =\left( 1-\frac{1}{\gamma }\right) +\frac{1}{%
\gamma }\frac{P(t_{i-1})}{P(t_{i})}= \\
& =1-\frac{1}{\gamma }\left( 1-\frac{P(t_{i-1})}{P(t_{i})}\right) =1-\frac{1%
}{\gamma }\frac{\Delta P}{P(t_{i})}  \notag
\end{align}
This relation is general \ for an ideal gas since it \ connects the final
volume to the initial volume for an irreversible adiabatic compression in
which the external pressure is suddenly increased.

It follows also that 
\begin{equation}
\frac{\Delta V(t_{i})}{V(t_{i-1})}=-\frac{1}{\gamma }\frac{\Delta P}{P(t_{i})%
}\;\;\;\;\;\;i.e.\;\;\;\;\;\;\frac{\Delta V}{V}\cong -\frac{1}{\gamma }\frac{%
\Delta P}{P}\;
\end{equation}
It is worth-while to observe that relation (35) coincides with the first
term of the expansion 
\begin{align*}
\frac{V(t_{i})}{V(t_{i-1})}& =\left( \frac{P(t_{i-1})}{P(t_{i})}\right) ^{%
\dfrac{1}{\gamma }}=\left( 1-\frac{P(t_{i})-P(t_{i-1}}{P(t_{i})}\right) ^{%
\dfrac{1}{\gamma }}= \\
& =1-\left( \frac{\Delta P}{P(t_{i})}\right) \frac{1}{\gamma }+\frac{1}{2!}%
\frac{1}{\gamma }\left( \frac{1}{\gamma }-1\right) \left( \frac{\Delta P}{%
P(t_{i})}\right) ^{2}+ \\
& +\frac{1}{3!}\frac{1}{\gamma }\left( \frac{1}{\gamma }-1\right) \left( 
\frac{1}{\gamma }-2\right) \left( \frac{\Delta P}{P(t_{i})}\right) ^{3}+....
\end{align*}
so for small $\Delta P$ steps we can evaluate $\Delta V$ from$%
\;\;P(t_{i})V(t_{i})^{\gamma }=$ $C.$

To evaluate $\Delta \mathcal{S}_{U}$ for an adiabatic step we need only $%
\Delta \mathcal{S}_{sys}.$ From the First Law we have 
\begin{align}
\Delta \mathcal{S}_{sys}(t_{i})& =\mathcal{S}_{sys}(t_{i})-\mathcal{S}%
_{sys}(t_{i-1})=\int_{t_{i-1}}^{t_{i}}\frac{\delta Q}{T}%
=\int_{t_{i-1}}^{t_{i}}\frac{C_{V}dT}{T}+\int_{t_{i-1}}^{t_{i}}\frac{PdV}{T}=
\\
& =C_{V}\;\ln \frac{T(t_{i})}{T(t_{i-1})}+R\;\ln \frac{V(t_{i})}{V(t_{i-1})}
\notag
\end{align}
which to the first order gives 
\begin{equation}
\Delta \mathcal{S}_{sys}(t_{i})=C_{V}\frac{T(t_{i})-T(t_{i-1})}{T(t_{i-1})}%
+R\left( -\frac{V(t_{i-1})-V(t_{i})}{V(t_{i-1})}\right)
\end{equation}
now, recalling that 
\begin{equation*}
\Delta T=-\Delta T(t_{i})=-\frac{P(t_{i})}{C_{V}}\Delta V(t_{i})
\end{equation*}
we have

\begin{align}
\Delta \mathcal{S}_{sys}(t_{i})& =-P(t_{i})\frac{V(t_{i})-V(t_{i-1})}{%
T(t_{i-1})}+R\left( -\frac{V(t_{i-1})-V(t_{i})}{V(t_{i-1})}\right) =  \notag
\\
& =-\frac{P(t_{i})}{P(t_{i-1})}R\frac{\Delta V}{V(t_{i-1})}+R\frac{\Delta V}{%
V(t_{i-1})}=R\frac{\Delta V}{V(t_{i-1})}\left( 1-\frac{P(t_{i})}{P(t_{i-1})}%
\right) =  \notag \\
& =R\frac{\Delta V}{V(t_{i-1})}\left( \frac{\gamma \Delta V}{%
V(t_{i-1})+\gamma \Delta V}\right) \cong R\gamma \left( \frac{\Delta V}{V}%
\right) ^{2}=\frac{R}{\gamma }\left( \frac{\Delta P}{P}\right) ^{2}
\end{align}
In the adiabatic expansion $\Delta P<0$ \ and $\Delta V>0,$ but these
changes do not alter the value of the entropic gain we have found in the
compression.

From relations (33) and (39) we can conclude that in the step-wise Carnot
cycle

\begin{equation}
\oint d\mathcal{S}_{U}\simeq \frac{\Delta P}{P}
\end{equation}
Since for each process \ the number of steps \ is $\mathcal{N}$ $\cong \frac{%
P}{\Delta P},$ and for each step $\Delta \mathcal{S}_{U}(t_{i})\sim \left(
\Delta P/P\right) ^{2}.$

The above estimate of $\Delta \mathcal{S}_{U}(t_{i})$ enables to write down
\ in explicit form the ``Dissipated Work''$\;\;W_{D}$: 
\begin{eqnarray*}
W_{D} &=&\sum_{i=1}^{2N}T(t_{i})\Delta \mathcal{S}_{U}(t_{i})=%
\sum_{i=1}^{n_{1}}T_{0}R\left( \frac{\Delta P}{P(t_{i})}\right)
^{2}+\sum_{i=1}^{n_{2}}T(t_{n_{1}+i})\frac{R}{\gamma }\left( \frac{\Delta P}{%
P(t_{n_{1}+i})}\right) ^{2}+ \\
&&+\sum_{i=1}^{n_{3}}T_{B}R\left( \frac{\Delta P}{P(t_{N+i})}\right)
^{2}+\sum_{i=1}^{n_{4}}T(t_{N+n_{3}+i})\frac{R}{\gamma }\left( \frac{\Delta P%
}{P(t_{N+n_{3}+i})}\right) .
\end{eqnarray*}


\begin{thebibliography}{9}
\bibitem{1}  To perform a true reversible process an infinity of steps are
needed. Here we consider a finite number of steps. As it is show in
Appendix, increasing the steps number we approximate better a reversible
cycle.

\bibitem{2}  E. Mach, ''Prinzipien der Warmelehre''-Leipzig (1896)

\bibitem{3}  A. Sommerfeld, ''Thermodynamics and Statistical
Mechanics''-Lectures on Theoretical Physics- Vol. V. Academic Press (1964)

\bibitem{4}  P. M. Morse,'' Thermal Physics''- Benjamin (1964)

\bibitem{5}  M.W. Zemansky, ''Heat and Thermodynamics'' MacGraw-Hill (1957)

\bibitem{6}  G. Trautteur (personal communication) pointed out that these
words remind those of Simone Weil in ''Reflexions \`{a} propos de la th\`{e}%
orie des quanta'' \ Les Cahiers du Sud, n$^{o}251,$(sign\`{e} Emile
Novis)(1942), in 'Sur la Science'- Gallimard (1966)

\bibitem{7}  It will be available at the following address \
http://physicsweb.org/TIPTOP/VLAB/
\end{thebibliography}
\end{document}